\documentclass[twocolumn]{aastex62}

\usepackage{amsmath}
\usepackage{amssymb}
\usepackage{placeins}

\graphicspath{{./}{figures/}}

\submitjournal{ApJ}

\shorttitle{The Contour Method}
\shortauthors{Goldstein \& Townsend}

\begin{document}


\title{The Contour Method: a new approach to finding modes of non-adiabatic stellar pulsations}


\correspondingauthor{Jacqueline Goldstein}
\email{jgoldstein@astro.wisc.edu}


\author{J. Goldstein}
\affil{Department of Astronomy, University of Wisconsin-Madison,
  2535 Sterling Hall, 475 N. Charter Street, Madison, WI 53706, USA}
\affil{Kavli Institute for Theoretical Physics, University of California, Santa Barbara, CA 93106, USA}

\author{R. H. D. Townsend }
\affil{Department of Astronomy, University of Wisconsin-Madison,
  2535 Sterling Hall, 475 N. Charter Street, Madison, WI 53706, USA}
\affil{Kavli Institute for Theoretical Physics, University of California, Santa Barbara, CA 93106, USA}



\newcommand{\Msun}{{\rm M}_{\odot}}
\newcommand{\Rsun}{{\rm R}_{\odot}}
\newcommand{\Lsun}{{\rm L}_{\odot}}

\newcommand{\Mstar}{M}
\newcommand{\Rstar}{R}
\newcommand{\Lstar}{L}

\newcommand{\bul}{\textbullet \hspace*{1mm}}
\newcommand{\dash}{\hspace*{2mm} \textendash}
\newcommand{\tab}{\hspace*{4mm}} 
\newcommand{\n}{\newline}

\newcommand{\Tth}{\tau_{\rm th}}
\newcommand{\Tdyn}{\tau_{\rm dyn}}

\newcommand{\by}{\boldsymbol{y}}
\newcommand{\bY}{\boldsymbol{Y}}
\newcommand{\bA}{\boldsymbol{A}}
\newcommand{\bBa}{\boldsymbol{B}^{\rm a}}
\newcommand{\bBb}{\boldsymbol{B}^{\rm b}}
\newcommand{\bI}{\boldsymbol{I}}
\newcommand{\bu}{\boldsymbol{u}}
\newcommand{\bS}{\boldsymbol{S}}


\newcommand{\mesa}{\textsc{mesa}}
\newcommand{\mesastar}{\textsc{mesastar}}
\newcommand{\gyre}{\textsc{gyre}}


\newcommand{\dy}{\mathrm{d}}
\newcommand{\yr}{\mathrm{yr}}
\newcommand{\kelv}{\mathrm{K}}
\newcommand{\kms}{\mathrm{km\,s^{-1}}}
\newcommand{\Teff}{T_{\rm eff}}


\newcommand{\med}{\operatorname{med}}
\newcommand{\real}{\operatorname{Re}}
\newcommand{\imag}{\operatorname{Im}}

\newcommand{\diff}{\mathrm{d}}
\newcommand{\ii}{\mathrm{i}}

\newcommand{\vxi}{\bmath{\xi}}
\newcommand{\vxih}{\vxi_{\rm h}}
\newcommand{\vv}{\mathbf{v}}

\newcommand{\vt}{v_{\theta}}
\newcommand{\vp}{v_{\phi}}

\newcommand{\Lrad}{L_{\rm rad}}

\newcommand{\xir}{\xi_{r}}
\newcommand{\xit}{\xi_{\theta}}
\newcommand{\xip}{\xi_{\phi}}

\newcommand{\txir}{\tilde{\xi}_{r}}
\newcommand{\txih}{\tilde{\xi}_{h}}
\newcommand{\tP}{\tilde{P}}
\newcommand{\trho}{\tilde{\rho}}
\newcommand{\tPhi}{\tilde{\Phi}}
\newcommand{\tS}{\tilde{S}}
\newcommand{\tT}{\tilde{T}}
\newcommand{\tf}{\tilde{f}}
\newcommand{\teps}{\tilde{\epsilon}}
\newcommand{\tkap}{\tilde{\kappa}}
\newcommand{\tLrad}{\tilde{L}_{\rm rad}}

\newcommand{\cP}{c_{P}}
\newcommand{\upsT}{\upsilon_{T}}
\newcommand{\nabad}{\nabla_{\rm ad}}

\newcommand{\kapad}{\kappa_{\rm ad}}
\newcommand{\kapS}{\kappa_{S}}
\newcommand{\epsad}{\epsilon_{\rm ad}}
\newcommand{\epsS}{\epsilon_{S}}
\newcommand{\crad}{c_{\rm rad}}
\newcommand{\dcrad}{\partial \crad}
\newcommand{\cdif}{c_{\rm dif}}
\newcommand{\cthm}{c_{\rm thm}}
\newcommand{\cepsS}{c_{\epsilon, S}}
\newcommand{\cepsad}{c_{\epsilon, {\rm ad}}}

\newcommand{\Yml}{Y^{m}_{\ell}}

\newcommand{\nablah}{\nabla_{\rm h}}

\newcommand{\sigmar}{\sigma_{\rm R}}
\newcommand{\sigmai}{\sigma_{\rm I}}
\newcommand{\sigmac}{\sigma_{\rm c}}

\newcommand{\omegar}{\omega_{\rm R}}
\newcommand{\omegac}{\omega_{\rm C}}
\newcommand{\omegai}{\omega_{\rm I}}

\newcommand{\Ds}{\mathcal{D}(\sigma)}

\newcommand{\Dw}{\mathcal{D}(\omega)}
\newcommand{\Dr}{\mathcal{D}_{\rm R}}
\newcommand{\Di}{\mathcal{D}_{\rm I}}
\newcommand{\Dc}{\mathcal{D}_{\rm C}}

\newcommand{\omeganad}{\omega_{\rm nad}}
\newcommand{\omegaad}{\omega_{\rm ad}}

\newcommand{\etamax}{\eta_{\rm max}}

\newcommand{\npg}{\tilde{n}}
\newcommand{\eps}{\varepsilon}

\newcommand{\Ocrit}{\Omega_{\rm crit}}
\newcommand{\veq}{v_{\rm eq}}
\newcommand{\veqstar}{v_{\rm eq,\ast}}

\newcommand{\nad}{NAd }
\newcommand{\ad}{Ad }
\newcommand{\itf}{ITF }

\newcommand{\needcitation}{{\color{blue}{citation }}}
\newcommand{\jg}[1]{\textcolor{red}{#1}}


\newcommand{\rtadd}[1]{\textcolor{brown}{#1}}
\newcommand{\rtdel}[1]{\textcolor{orange}{\sout{#1}}}

\begin{abstract}

The contour method is a new approach to calculating the non-adiabatic pulsation frequencies of
stars. These frequencies can be found by
solving for the complex roots of a characteristic equation constructed from the linear non-adiabatic stellar pulsation equations. A complex-root solver requires an initial trial frequency for each non-adiabatic root. A standard method for obtaining initial trial frequencies is to use a star's adiabatic pulsation frequencies, but this method can fail
to converge to non-adiabatic roots, especially as  the growth and/or damping rate of the pulsations becomes large. The contour method provides an alternative way for obtaining initial trial frequencies that robustly
converges to non-adiabatic roots, even for stellar models with extremely non-adiabatic pulsations and thus large growth/damping rates. We describe the contour method implemented in the \gyre\ stellar pulsation code and use it to calculate the non-adiabatic pulsation frequencies of $10\,\Msun$ and $20\,\Msun$ $\beta$ Cephei  star models, and of a $0.9\,\Msun$ extreme helium star model.

\end{abstract}




\section{Introduction}
\label{sec:introduction}

Stars across the Hertzsprung-Russel diagram (HRD) exhibit pulsations that carry information about stellar structure and evolution. Modeling stellar pulsations requires solving the stellar pulsation equations \citep[e.g.][]{1989_Unno, 2010_Aerts_etal} as a boundary eigenvalue problem, to obtain eigenfrequencies and eigenfunctions. Many stellar pulsation codes solve for eigenfrequencies by finding the roots of a characteristic equation,
\begin{align}
\Ds = 0,
\label{eq:Dw}
\end{align}
where $\Ds$ is a discriminant function, and $\sigma$ is the pulsation angular frequency.
Pulsation codes approach constructing discriminant functions in different ways, and even the same code can implement a variety of approaches; but the roots should agree within and across codes because they represent the intrinsic eigenfrequencies of the star.

How pulsation codes construct and solve the
characteristic equation depends on whether the pulsations being modeled are adiabatic or non-adiabatic. For adiabatic pulsations, the
linear adiabatic (LA) stellar pulsation equations yield $\Ds$ and $\sigma$ that are real-valued, and the roots are guaranteed to be found through standard bracketing approaches such as bisection \citep[e.g.,][]{1992_Press}.

For non-adiabatic pulsations, the linear non-adiabatic (LNA) stellar pulsation equations yield $\Ds$ and $\sigma$ that are complex-valued. The pulsation frequency can be written
\begin{equation}
  \sigma = \sigmar + \ii \sigmai,
\end{equation}
where `R' and `I' denote real and imaginary parts, respectively. Assuming pulsations have a time dependence $\propto \exp(-\ii \sigma t)$, $\sigmar$ describes the oscillatory behavior of the pulsation, while $\sigmai$ describes overstable growth ($\sigmai > 0$) or damped decay ($\sigmai < 0$). 

Complex roots cannot meaningfully be bracketed; therefore, solving the characteristic equation~(\ref{eq:Dw}) in the LNA case requires iterative improvement of an initial trial frequency using, for instance, the Newton-Raphson or secant algorithms \citep[e.g.,][]{1992_Press}. These complex-root solvers share the disadvantage that convergence is only guaranteed when the trial frequency is sufficiently close to a root. Challenges arise when pulsations become increasingly non-adiabatic because the trial frequencies (e.g., established from adiabatic eigenfrequencies, see Sec.~\ref{sec:background}) can be distant from the roots; consequently, the solver converges to the wrong root or does not converge at all. The result is an incorrect or incomplete non-adiabatic pulsation analysis.

To address this problem, we describe and apply a new \textit{contour method} for generating initial trial frequencies. The contour method has two main benefits over other approaches. First, it successfully finds all non-adiabatic pulsation frequencies for tested stellar models and frequency ranges. Second, it generates a `contour map' that can be used to visualize the global non-adiabatic pulsation properties of a stellar model.

In Sec. \ref{sec:background} we review two approaches used by existing stellar pulsation codes to generate initial trial frequencies.
In Sec. \ref{sec:contour} we introduce the contour method and describe its implementation in the \gyre\ stellar pulsation code.
In Sec.~\ref{sec:calculations} we compare these various methods in calculating the non-adiabatic pulsation frequencies of three stellar models: $10\,\Msun$ and $20\,\Msun$ $\beta$ Cephei stars, and a $0.9\,\Msun$ extreme helium star. We show that the contour method finds non-adiabatic pulsation frequencies missed by other methods.
In Sec.~\ref{sec:discussion} we address the computational cost of the contour method, and discuss ways that it can be mitigated. The contour method will be available in release 6.0 of the \gyre\ code, providing a new tool for modeling the unprecedented observational stellar pulsation data collected by the Transiting Exoplanet Survey Satellite \citep[\emph{TESS};][]{2014_Ricker} and other future missions.


\section{Background}
\label{sec:background}

\subsection{Methods for Obtaining Initial Trial Frequencies}

The most common approach for generating initial trial frequencies, which we call the \textit{adiabatic method}, is to first solve the LA stellar pulsation equations. This method is first described by \cite{1971_Castor}, who constructs an adiabatic $\Ds$ and solves for its real-valued roots. The roots are perturbed, resulting in quasi-adiabatic frequencies that are used as initial trials for the complex roots of a corresponding non-adiabatic $\Ds$. Similar methods, using the unperturbed adiabatic roots, are implemented in the
\textsc{boojum} \citep{2005a_Townsend}, \textsc{lnawenr}
\citep{2008_Suran}, and \gyre\ \citep{2013_Townsend,2018_Townsend}
non-adiabatic pulsation codes.

The adiabatic method, however, has a weakness. As a pulsation becomes increasingly non-adiabatic, that is as the imaginary component of the frequency, $\sigmai$, increases in magnitude, the real component of the frequency, $\sigmar$, typically shifts away from the adiabatic frequency.  As a result, the non-adiabatic frequencies can interlace the adiabatic ones. Consequently, when adiabatic roots are used as initial trial frequencies, the complex-root solver can converge to the non-adiabatic roots of neighboring modes, missing modes in the process. If pulsations are extremely non-adiabatic, that is $|\sigmai / \sigmar| \gtrsim 1 $, the non-adiabatic roots can be far enough from the adiabatic ones that the complex-root solver doesn't converge at all (see Sec. \ref{sec:calculations}).

Another approach to generating initial trial frequencies, which we call the \textit{minimum modulus method}, was proposed by \cite{1990a_Gautschy}. They construct a non-adiabatic $\Ds$ and evaluate its modulus, $|\Ds|$, as a function of $\sigmar$ to look for minima along the real axis ($\sigmai=0$). The values of $\sigmar$ at these minima then serve as initial trials for the complex roots of $\Ds$.

The minimum modulus method, however, also has weaknesses. The first, similar to the adiabatic method, is that if pulsations are extremely non-adiabatic, some of the roots of $\Ds$ may be so far from the real axis that the complex-root solver doesn't converge (see Sec. \ref{sec:calculations}). The second weakness is that because the method relies on the modulus of a complex function, there is a degeneracy of roots that are complex conjugates. This was shown to occur in an extreme limit of non-adiabaticity associated with the strange instability \citep{1990b_Gautschy}.


\section{The Contour Method}
\label{sec:contour}
\begin{figure}
	\centering \includegraphics[width=\linewidth]{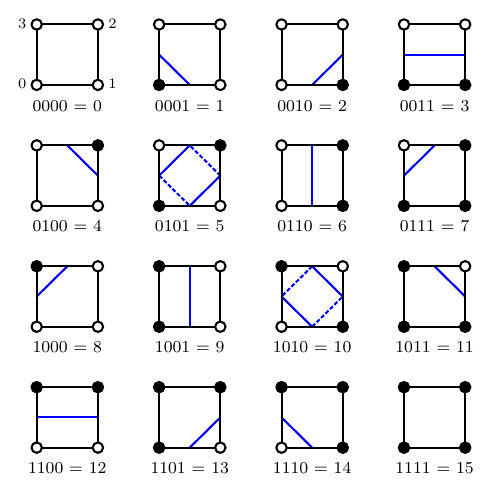}
		\caption{Look-up table for the marching  squares algorithm, showing the 16 possible configurations that can arise, and labeled by their index (in binary and decimal). Cell corners are plotted as circles; filled if the discriminant component ($\Dr$ or $\Di$) is positive at that corner, and open if it is negative. Configuration 0 (top-left) shows the labels $i=0,\ldots,3$ for each corner.  The blue lines show example linear contour segments corresponding to each configuration. For configurations 5 and 10, there are two possible pairs of segments, shown using solid and dotted lines; \gyre\ adopts the pair with the shorter total length.}  
\label{fig:marching_squares}
\end{figure}
\begin{figure}
	\centering
	\includegraphics[width=\linewidth]{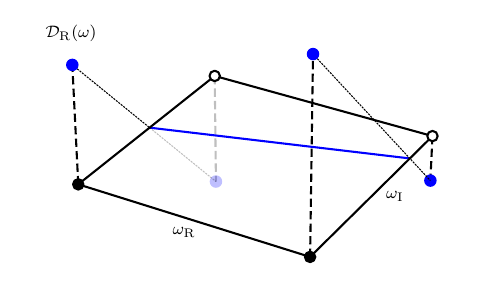}
	\caption{The linear interpolation process used to approximate where contour segments connect to cell edges. Illustrated here is the discriminant component $\Dr$ for a cell with configuration 3 (see Fig. \ref{fig:marching_squares}).}
\label{fig:marching_squares_segment}
\end{figure}
\begin{figure}
	\centering
	\includegraphics[width=\linewidth]{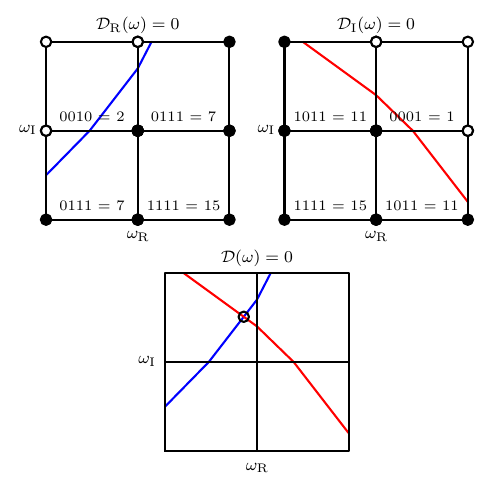}
	\caption{\textbf{Top:} An example grid showing cells labeled by their configuration index and corresponding zero-contour segments for discriminant component $\Dr$ (left) and $\Di$ (right).
	\textbf{Bottom:} Example contour intersection for $\Dw$. The point in the top-left cell where the segments intersect (highlighted with a circle) is an approximate root of $\Dw$. The intersection serves as an initial trial frequency for the complex-root solver.}
	\label{fig:marching_squares_intersect}
\end{figure}

In the $\gyre$ code, the contour method works by calculating a complex-discriminant function, $\Dw$, on a grid in the complex-$\omega$ plane\footnote{We note that the contour method performs equally well using the dimensioned angular frequency $\sigma$ in place of $\omega$; however, most pulsation codes, including \gyre, work internally with $\omega$.}. Here, $\omega$ is the dimensionless frequency, defined by
\begin{equation}
  \omega = \sqrt{\frac{\Rstar^3}{G \Mstar}} \sigma,
\end{equation}
where $\Mstar$ is the stellar mass and $\Rstar$ the stellar radius. This grid is then used to interpolate the zero-contours of the real and imaginary components of the discriminant, $\Dr$ and $\Di$, respectively. The intersections between real and imaginary zero-contours approximate the roots of $\Dw$ and serve as initial trial frequencies for the complex-root solver.

\subsection{Constructing the Contours}

We implement the contour method using the ‘marching squares’ algorithm \citep[see, e.g.,][]{2013_Wenger}, which generates zero-contours for a two-dimensional scalar field on a grid.  First, a rectangular grid with a user-specified range and resolution in the complex-$\omega$ plane is constructed, and $\Dw$ is evaluated at each grid point. This step can be computationally expensive, but it is ideally suited to parallel execution across a distributed cluster (see Sec.~\ref{sec:discussion} for further discussion).

The zero-contours are then constructed by considering each rectangular cell defined by four adjacent grid points. These corner points are labeled in counter-clockwise order with an integer $i$, starting from $i=0$ in the cell's lower-left corner. Each corner is assigned a value based on the sign of the discriminant component ($\Dr$ or $\Di$) at its location: $2^{i}$ if the component is positive, and 0 if it is negative. The 
values for each corner are summed to determine a configuration index 0--15 for the cell. 

This index is used to access a look-up table, illustrated in Fig.~\ref{fig:marching_squares}, that specifies which cell edges should be connected by linear contour segments separating negative and positive corners. Cells with all positive corners (configuration 0) and with all negative corners (configuration 15) don't have any segments within them. Cells with diagonal pairs of negative and positive corners (configurations 5 and 10) have two possible pairs of contour segments. \gyre\ chooses 
the pair with the shorter total length,
but the degeneracy can be broken by constructing a higher resolution grid.

The location where a contour segment connects to a cell edge is approximated by a linear interpolation between the discriminant component values at the two corners. We illustrate this in Fig.~\ref{fig:marching_squares_segment} for a cell with configuration 3.

\subsection{Contour Intersections as Initial Trial Frequencies}

In cells containing zero-contour segments of both $\Dr$ and $\Di$,
$\gyre$ determines whether the segments intersect within the cell.
If so, the intersection approximates where $\Dw=0$, and is adopted as an initial trial frequency for the complex-root solver. Fig.~\ref{fig:marching_squares_intersect} illustrates this process.

A powerful feature of the contour method is that, when combined across cells, the zero-contour segments build a contour map that provides a rich visual representation of the global pulsation properties of a model across a given frequency
range. We illustrate contour maps in the following section.


\section{Calculations}
\label{sec:calculations}

In this section we compare and contrast the various methods for
generating initial trial frequencies (Sections~\ref{sec:background}
and~\ref{sec:contour}) in the context of $\beta$ Cephei stars and
extreme helium (EHe) stars.

$\beta$ Cephei stars \citep[e.g.][]{2005_Stankov} are main sequence
stars with masses $M \gtrsim 8\,\Msun$ that exhibit low order pressure and gravity modes driven by the iron-bump $\kappa$ mechanism \citep{1992_Cox, 1993a_Dziembowski}. The pulsations are weakly non-adiabatic, but increase in non-adiabaticity
toward higher frequencies and higher masses.

EHe stars \citep[e.g.][]{2008a_Jeffery} are rare, low mass, high
luminosity, early-type supergiants that belong to a class of
hydrogen-deficient carbon stars. It remains an open question how these
stars became depleted of their hydrogen. EHe stars exhibit
pressure modes and strange modes, driven by both the helium
$\kappa$ mechanism and by the strange-mode instability, which occurs in the presence of extreme non-adiabaticity.

We use release 12778 of the \textsc{mesa} stellar evolution code
\citep{2011_Paxton, 2013_Paxton, 2015_Paxton, 2018_Paxton, 2019_Paxton} to construct models for $10\,\Msun$ and $20\,\Msun$ $\beta$ Cephei stars, and for a 0.9$\Msun$ EHe star; we describe these models in the following sections. We then apply \gyre\ with the different methods for generating initial trial frequencies to compare the resulting non-adiabatic pulsation analyses. 

\pagebreak

\subsection{$10\,\Msun$ $\beta$ Cephei Star Model}

The $10\,\Msun$ stellar model is evolved from zero-age main-sequence (ZAMS) to the terminal-age main-sequence (TAMS), when the core hydrogen mass fraction, $X_{\rm c}$, drops below $10^{-5}$. OPAL opacity tables are used with the proto-solar initial abundances from \citet{2009_Asplund}, and we neglect any rotation or mass loss. Convection is modeled with a mixing-length parameter $\alpha_{\rm MLT} = 1.8$ but no overshoot, and convective boundaries are determined using the predictive mixing scheme described in \citet{2018_Paxton} with the Ledoux stability criterion.

\begin{figure}[!htb]
	\centering
	 \includegraphics[width=\linewidth,trim=0.1in 0.2in 0in 0.1in,clip]{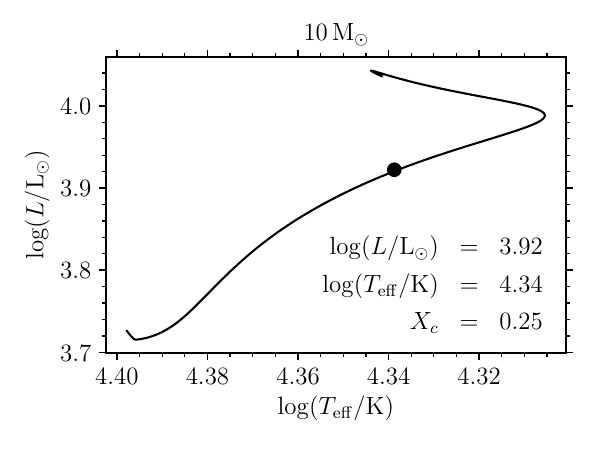}
	 \caption{ HRD showing the evolutionary track for the
             $10\,\Msun$ $\beta$ Cephei star model. The snapshot considered
             in the text is indicated by the filled circle, and
             labeled with its stellar parameters (luminosity, $L$; effective temperature, $\Teff$; core
             hydrogen mass fraction, $X_{\rm c}$).}
     \label{fig:10_0_hr_diag}
\end{figure}

We focus on a specific snapshot of the model chosen with $X_{\rm c}=0.25$, which places it well inside the $\beta$ Cephei instability strip for radial modes \citep[e.g.][]{2015_Paxton}. The parameters of this snapshot, and its position in the HRD, are shown in Fig.~\ref{fig:10_0_hr_diag} along with the star's main-sequence evolutionary track.

\subsubsection{Adiabatic Roots as Initial Trial Frequencies}

\begin{figure*}[!htb]
    \centering \includegraphics*[width=\linewidth,trim=0.1in 0.2in 0in
      0.1in,clip]{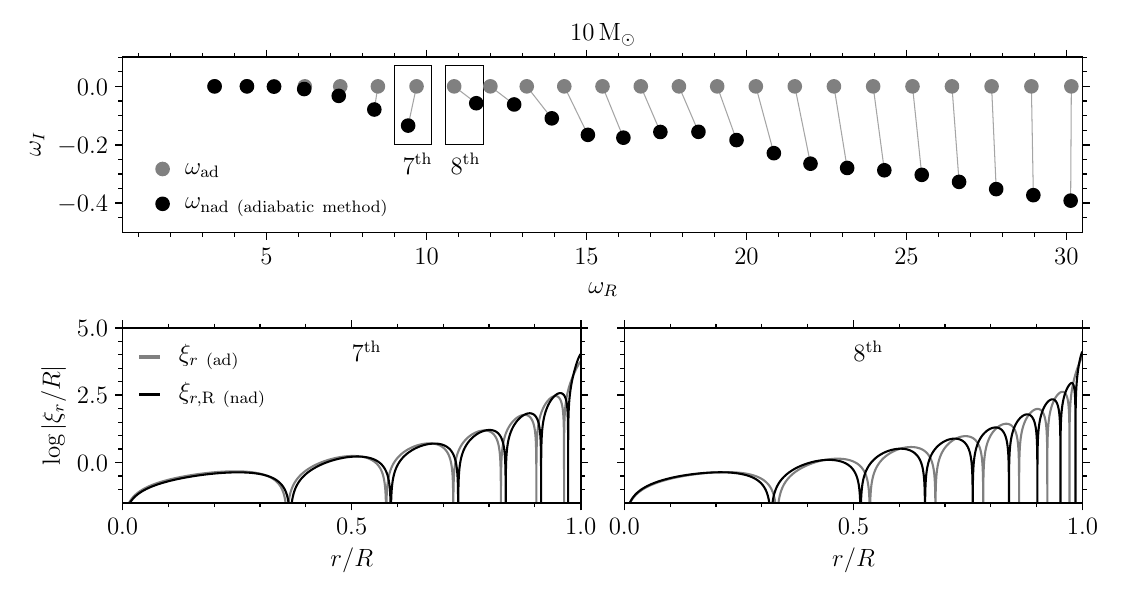}
    \caption{\textbf{Top:} Dimensionless frequencies of radial
	    	modes in the complex-$\omega$ plane, found using the adiabatic method for the $10\,\Msun$ snapshot marked in Fig.~\ref{fig:10_0_hr_diag}.  Lines join each non-adiabatic frequency ($\omeganad$, black filled circles) to the adiabatic frequency ($\omegaad$, grey filled circles) that was used as the initial trial frequency for the complex-root solver. 
	    	\textbf{Bottom:} The dimensionless radial displacement wave  functions, $\xi_r/R$, plotted as a function of fractional radius, $r/R$, for the frequencies boxed in the top panel (adiabatic, grey; non-adiabatic, black). In the non-adiabatic cases, we show only the real part of the wave function,  $\xi_{r,{\rm R}}$. The wave functions for the $7^{th}$ boxed frequency pair, shown on the left, exhibit 7 radial nodes for both adiabatic and non-adiabatic cases. The wave functions for the $8^{th}$ boxed frequency pair, shown on the right,  exhibit 8 radial nodes in the adiabatic case, but 9 in the non-adiabatic case. The non-adiabatic mode with 8 nodes is missing.}
    \label{fig:10_0_ad_nad_xi}
\end{figure*}

We search for non-adiabatic radial modes of the $10\,\Msun$ snapshot, using adiabatic frequencies in the range $0.5 \leq \omegar \leq 30.5$ as initial trial frequencies (the adiabatic method; see
Section~\ref{sec:background}). In the top panel of Fig.~\ref{fig:10_0_ad_nad_xi} we show the adiabatic and non-adiabatic dimensionless pulsation frequencies, $\omegaad$ and $\omeganad$ respectively, in the complex-$\omega$ plane. Lines join each non-adiabatic frequency back to the adiabatic frequency that was used as its initial trial frequency.

Not immediately apparent in the figure is the fact that one of the non-adiabatic modes is missing. We see this when we
examine the radial wave functions of modes with consecutive frequencies, expecting the wave functions to exhibit
likewise-consecutive numbers of radial nodes.\footnote{Strictly, this consecutive node numbering property applies only to solutions of the radial LA equations, which are of regular Sturm-Liouville form \citep[e.g.,][]{1958_Ledoux}. However, in the present case the radial modes are only modestly non-adiabatic, and so the property should also apply to the solutions of the radial LNA equations.}

In the bottom panel of Fig.~\ref{fig:10_0_ad_nad_xi} we plot the dimensionless radial displacement wave functions of the consecutive frequency pairs boxed in the top panel. In the non-adiabatic cases, we show only the real part of the wave function, $\xi_{r,{\rm R}}$. For the 7$^{\rm th}$ frequency pair, the adiabatic and non-adiabatic wave functions both show 7 nodes, as we expect. For the 8$^{\rm th}$ frequency pair, however, the adiabatic wave function shows 8 radial nodes, but the associated non-adiabatic wave function shows 9. The non-adiabatic mode with 8
radial nodes is missing. This illustrates the problem with using adiabatic frequencies as initial trial frequencies; the root solver does not always converge to the correct non-adiabatic ones.

\begin{figure*}[!p]
    \centering \includegraphics[width=\linewidth,trim=0.1in 0.2in 0in
      0.1in,clip]{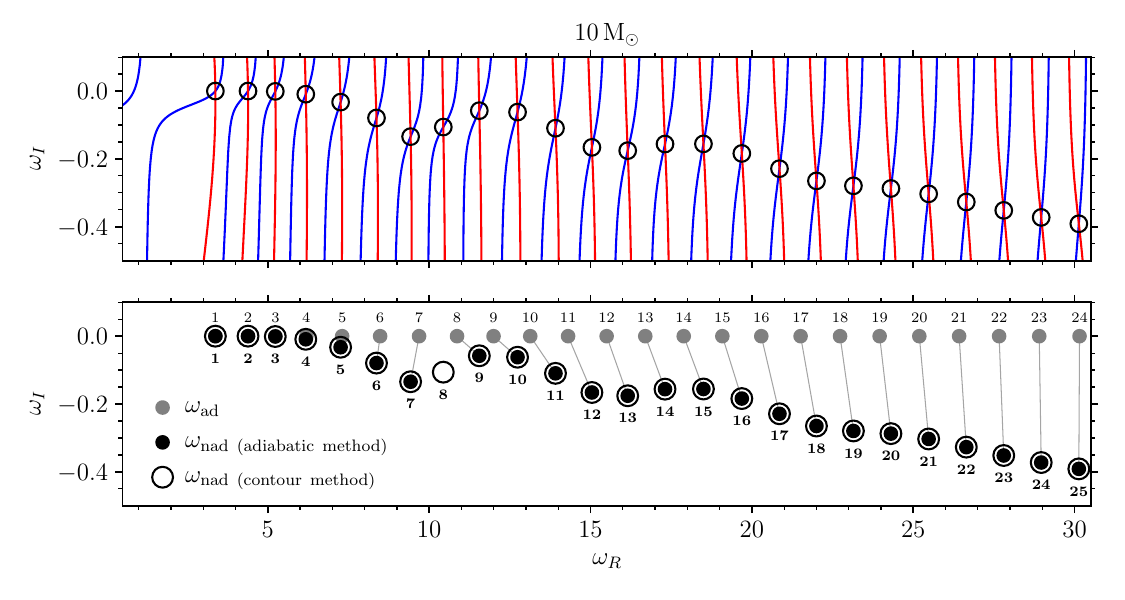}
    \caption{ \textbf{Top:} Contour map showing the
  		    zero-contours of the real ($\Dr$, blue) and imaginary ($\Di$, red) components of the discriminant function in the complex-$\omega$ plane, for the
      		$10\,\Msun$  snapshot marked in Fig.~\ref{fig:10_0_hr_diag}. The intersections are approximate roots of $\Dw$ and serve as initial trial frequencies for the complex-root solver. The open circles indicate the roots actually found by the solver.  
      		\textbf{Bottom:} Dimensionless
      		non-adiabatic frequencies of radial modes found using the contour method (open circles) overlain on those found using the adiabatic method (grey and filled black circles, taken from the top panel of Fig.~\ref{fig:10_0_ad_nad_xi}). Each mode is labeled by the number of radial nodes in its radial displacement wave function. Note how the $8^{\rm th}$ non-adiabatic mode was missed by the adiabatic method.}
    \label{fig:10_0_ad_nad_map}
    
    \includegraphics[width=\linewidth,trim=0.1in 0.2in 0in
      0.1in,clip]{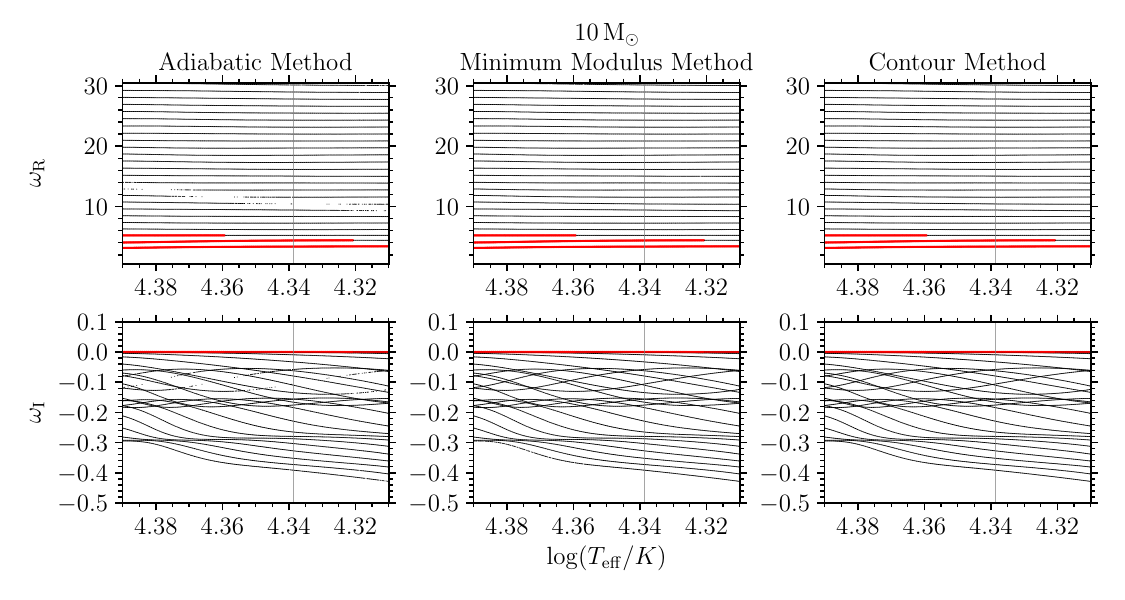}
    \caption{Modal diagrams showing the radial non-adiabatic
    		 frequencies (stable, black; overstable, red) of radial
    	     modes for the $10\,\Msun$ model as it evolves
             across the main sequence. \textbf{Top:} The real part of the dimensionless pulsation frequency $\omegar$, as a function of effective temperature, $\Teff$; 
      		\textbf{Bottom:} The corresponding imaginary part,
    	    $\omegai$. The diagrams are constructed
  		 	  using the adiabatic method (left), minimum modulus method
  		    (middle), and contour method (right). The $10\,\Msun$
    		snapshot shown in Figs.~\ref{fig:10_0_hr_diag}--\ref{fig:10_0_ad_nad_map} is indicated by a vertical grey line.}
    \label{fig:10_0_modal_diag}
\end{figure*}

To see which modes are missed as the $10\,\Msun$ model evolves across the main sequence, we repeat our calculations for each timestep  between the ZAMS and the TAMS. The left column of Fig.~\ref{fig:10_0_modal_diag} shows a \textit{modal diagram} constructed from these calculations, plotting the non-adiabatic frequencies (upper-panel: $\omegar$, lower-panel: $\omegai$) of radial modes as a function of the effective temperature $\Teff$. To improve the clarity of this and other modal diagrams, we only show models monotonically decreasing in effective temperature, neglecting the Henyey hook portion near the TAMS  when the star evolves to the blue in the HRD. Overstable modes are marked in red. A band of frequencies is missing, indicating where (as in Fig.~\ref{fig:10_0_ad_nad_xi}) the complex-root solver converged to the wrong non-adiabatic frequencies.

\subsubsection{Minimum Modulus as Initial Trial Frequencies}

For comparison, we repeat our calculations for the $10\,\Msun$ model using the minimum modulus method (Sec. \ref{sec:background}) implemented in $\gyre$. We show the resulting modal diagram in the middle column panel of Fig.~\ref{fig:10_0_modal_diag}. The minimum
modulus method fills in the frequencies that were missed using the adiabatic method. This is because the non-adiabatic frequencies are close to the $\omegar$ axis, and therefore produce well-defined minima in $|\Dw|$
along this axis.

\subsubsection{Contour Intersections as Initial Trial Frequencies}

We again repeat our calculations for the $10\,\Msun$ model, now using the contour method (Sec. \ref{sec:contour}).  We use a grid of 1000 points spanning $0.5 \leq \omegar \leq 30.5$, and 400 points spanning $-6 \leq \omegai \leq 6$, so that the grid spacing is the same in both dimensions. We show the contour map in the top panel of  Fig.~\ref{fig:10_0_ad_nad_map}, displaying the zero-contours of $\Dr$ and $\Di$. The intersections of the contours are used as initial trial frequencies for the complex-root solver.

In the bottom panel of Fig.~\ref{fig:10_0_ad_nad_map}, we
compare the modes found using the contour method with those found using the adiabatic method, shown in the top panel of Fig.~\ref{fig:10_0_ad_nad_xi}. The contour method recovers all the modes previously found,  but also finds the missing mode with 8 radial nodes. We now see that with the adiabatic method, the 8$^{\rm th}$ adiabatic frequency converged to the 9$^{\rm th}$ non-adiabatic frequency. Each subsequent adiabatic frequency converged to the wrong non-adiabatic one. This highlights the problem with using adiabatic frequencies as initial trial frequencies  even for weakly non-adiabatic pulsation. The contour method, on the other hand, provides initial trial frequencies that are close to the true roots, resulting in the robust convergence to all non-adiabatic frequencies.

We show the modal diagram for the contour method in the right column of Fig.~\ref{fig:10_0_modal_diag}. The contour method fills in the frequencies that were previously missed when using the adiabatic method. There is no difference between the modal diagrams for the minimum modulus and contour methods here for the $10\,\Msun$ model, but --- as we shall demonstrate --- the contour method still succeeds when the pulsations become strongly non-adiabatic and the other methods fail.

\subsection{$20\,\Msun$ $\beta$ Cephei Star Model}

\begin{figure}[!htb]
	\centering		
	\includegraphics[width=\linewidth,trim=0.1in 0.2in 0in 0.1in,clip]{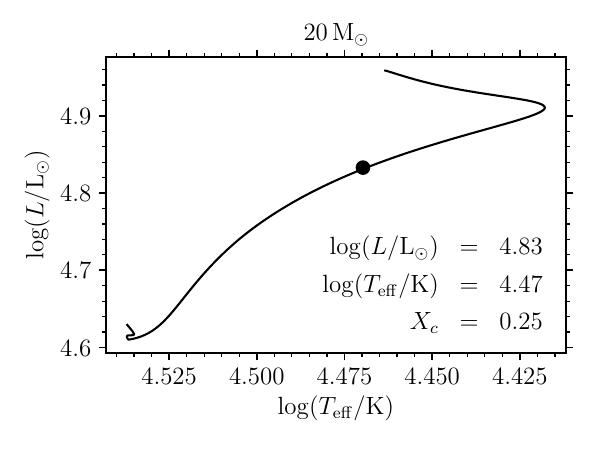}
    \caption{HRD showing the evolutionary
      track for the $20\,\Msun$ $\beta$ Cephei star model. The snapshot
      considered in the text is indicated by the filled circle, and
      labeled with its stellar parameters (luminosity, $L$;
      effective temperature, $\Teff$; core hydrogen mass fraction, $X_{\rm c}$).}
      \label{fig:20_0_hr_diag}
\end{figure}

We now repeat our analysis for a more massive $20\,\Msun$ stellar model, calculated in the same manner as the $10\,\Msun$ model. We begin by focusing on a snapshot chosen with $X_{\rm c}=0.25$, as before, marked in Fig.~\ref{fig:20_0_hr_diag}. Due to its larger luminosity-to-mass ratio, we expect the pulsations of this model to be more non-adiabatic than the $10\,\Msun$ case \citep[see, e.g.,][]{1984_Saio}. 

\begin{figure*}[!htb]
	\centering
	\includegraphics[width=\linewidth,trim=0.1in 0.2in 0in 0.1in,clip]{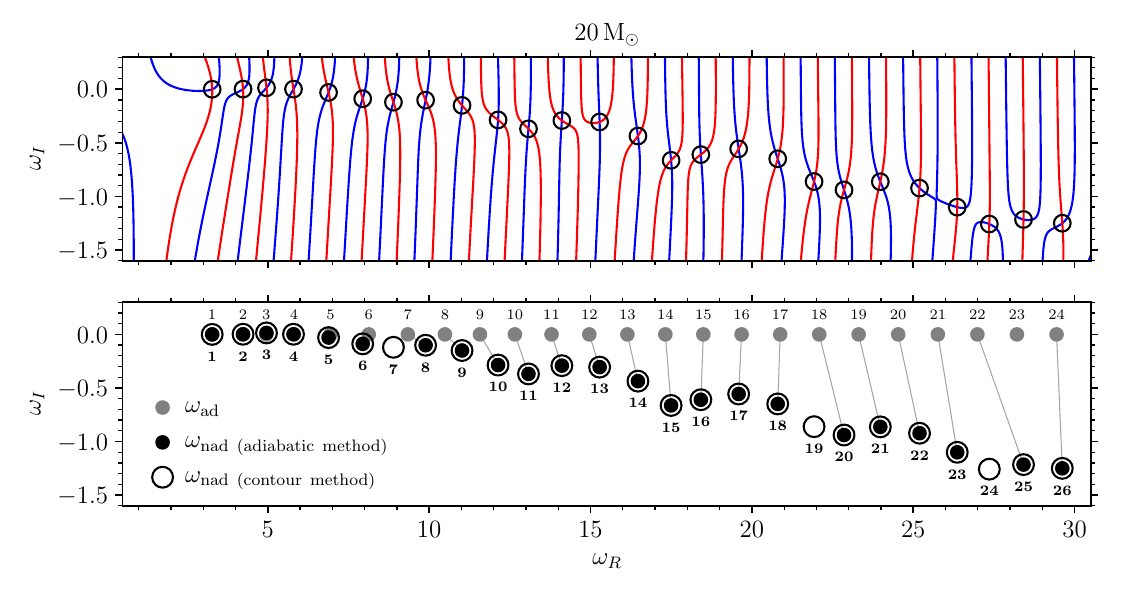}
	\caption{ As in Fig.~\ref{fig:10_0_ad_nad_map}, except the $20\,\Msun$
             snapshot marked in Fig.~\ref{fig:20_0_hr_diag} is shown. Note how three modes are now missed by
            the adiabatic method.}
	\label{fig:20_0_ad_nad_map}

	\includegraphics[width=\linewidth,trim=0.1in 0.2in 0in
          0.1in,clip]{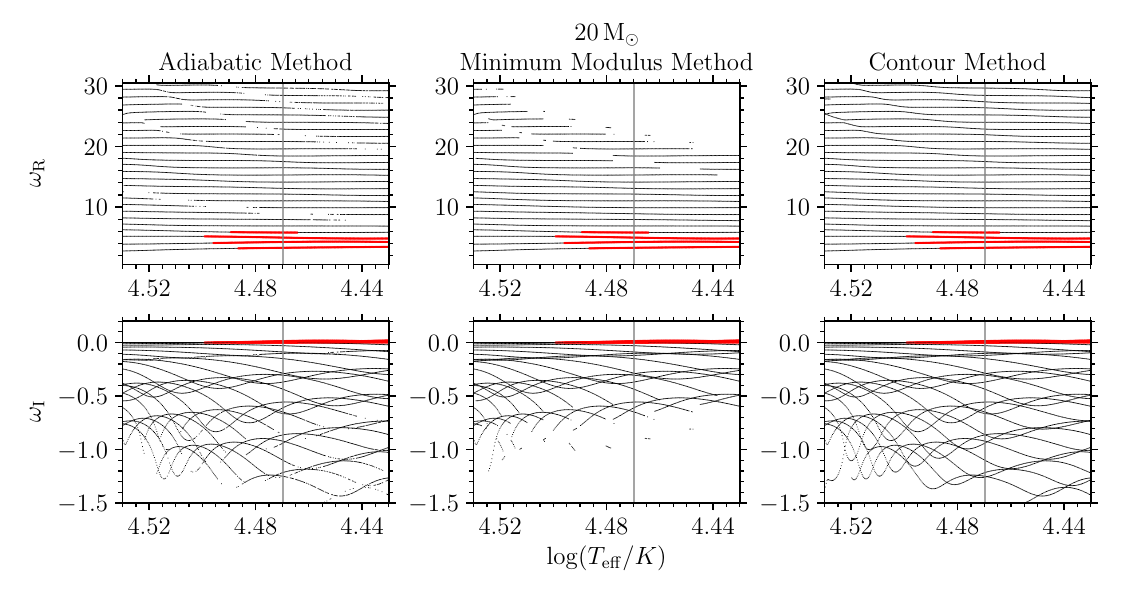}
	\caption{As in Fig.~\ref{fig:10_0_modal_diag}, except the $20\,\Msun$
			model is shown.}
		\label{fig:20_0_modal_diag}
\end{figure*}

In the top panel of Fig.~\ref{fig:20_0_ad_nad_map} we show the contour
map for the $20\,\Msun$ snapshot, along with the intersections that
are used as initial trial frequencies. In the bottom panel of
Fig.~\ref{fig:20_0_ad_nad_map}, we compare the non-adiabatic
frequencies found using the contour method with those found using the
adiabatic method. The figure reveals that three non-adiabatic frequencies are missed using the adiabatic method.

In Fig.~\ref{fig:20_0_modal_diag} we show the modal diagrams
for the $20\,\Msun$ model constructed using the adiabatic
method (left), minimum modulus method (middle), and the contour method (right).  Multiple bands of frequencies are missed by the adiabatic method toward larger $\omegar$; the missing frequencies are apparently coincident with avoided crossings. The minimum modulus method also experiences difficulties near avoided crossings, but it is also unable to find non-adiabatic frequencies with $|\omegai| \gtrsim 0.75$; this is because the minima in $|\Dw|$ disappear when roots become too distant from the $\omegar$ axis. Only the contour method finds all the non-adiabatic frequencies, as can be seen from the complete modal diagram.


\subsection{Extreme Helium Star Model}

\begin{figure}[!htb]
	\centering
	\includegraphics[width=\linewidth,trim=0.1in 0.2in 0in 0.1in,clip]{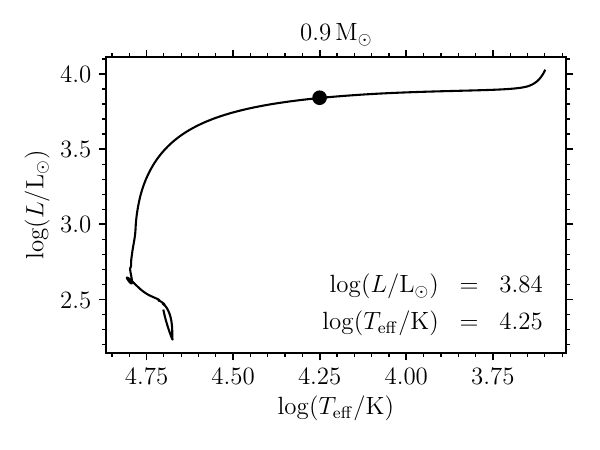}
	\caption{HRD showing the evolutionary track for the $0.9\,\Msun$ 
		extreme helium star model. The snapshot considered in the text is indicated by the
  		filled circle, and labeled with its stellar parameters (luminosity, $L$, and effective temperature, $\Teff$).}
	\label{fig:0_9_hr_diag}
\end{figure}

The contour method is especially powerful for studying extremely non-adiabatic pulsations with large  growth/damping rates ($|\omegai / \omegar| \gtrsim 1$). To demonstrate this, we repeat our analysis for a $0.9\,\Msun$ EHe star model constructed to be qualitatively similar to the case studied by \citet[][see their Fig.~1]{1990b_Gautschy}. The model is created at the He-ZAMS with an initially uniform composition given by the mass fractions $X=0$, $Y=0.903$, $Z=0.097$ and the R2 abundance profile from \citet{1987_Weiss}; other modeling parameters are the same as for the $\beta$ Cephei star models. It is evolved post He-TAMS until it reaches an effective temperature $\log (\Teff/{\rm K}) = 3.6$. We first focus on the snapshot of the model chosen about half-way along its trip to the red, $\log (\Teff/{\rm K}) = 4.25$. The parameters of this snapshot, and its position in the HRD, are shown in Fig.~\ref{fig:0_9_hr_diag} along with the star's evolutionary track. 

\begin{figure*}[!htb]
\centering \includegraphics[width=\linewidth,trim=0.1in 0.2in 0in
  0.1in,clip]{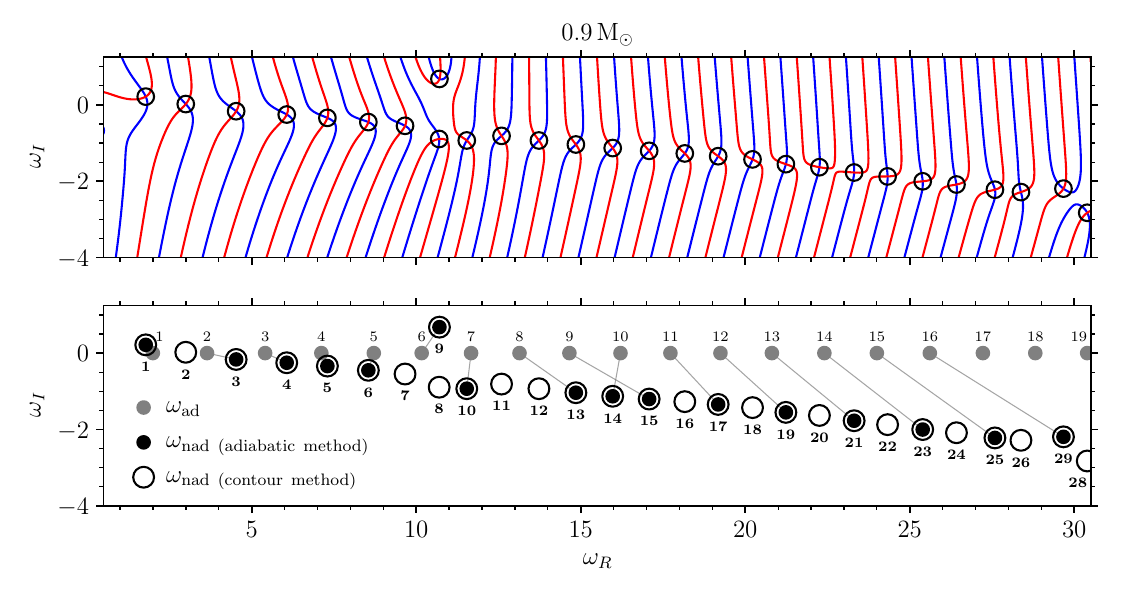}
	\caption{As in Fig.~\ref{fig:10_0_ad_nad_map}, except the $0.9\,\Msun$
            snapshot marked in Fig.~\ref{fig:0_9_hr_diag} is shown. Note the increasing number of non-adiabatic modes missed by the adiabatic method, and that there is no non-adiabatic mode with 27 radial nodes.}
	\label{fig:0_9_ad_nad_map}

	\includegraphics[width=\linewidth,trim=0.1in 0.2in 0in 0.1in,clip]{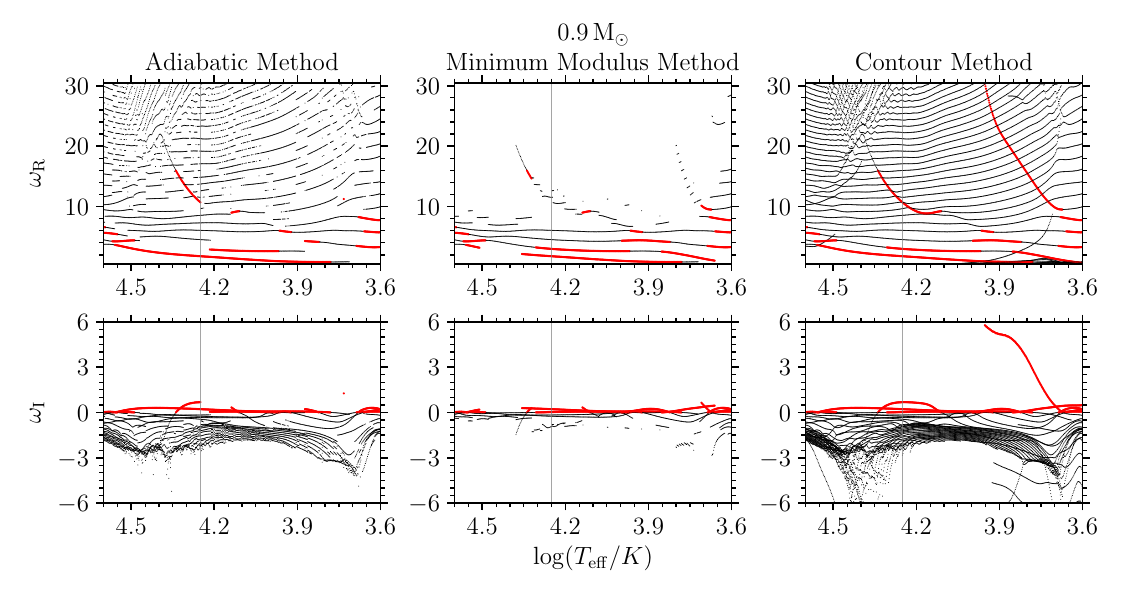}
	\caption{As in Fig.~\ref{fig:10_0_modal_diag}, except the $0.9\,\Msun$ 
            model is shown.}
	\label{fig:0_9_modal_diag}
\end{figure*}

In the top panel of Fig.~\ref{fig:0_9_ad_nad_map} we show the contour map for the $0.9\,\Msun$ snapshot along with the intersections used as initial trial frequencies. In the bottom panel we compare the frequencies found using the contour method with those using the adiabatic method. The modes are more non-adiabatic compared with the $\beta$ Cephei star models, and
many frequencies are missed by the adiabatic method. The contour method, in contrast, robustly finds all frequencies. We see that there is no non-adiabatic mode with 27 radial nodes, which is a physical effect due to non-adiabaticity captured by the contour method, and not the result of a mode missed by the method.

In Fig.~\ref{fig:0_9_modal_diag} we show the modal diagrams for
 the $0.9\,\Msun$ model, constructed using the adiabatic method
(left), minimum modulus method (middle), and contour method (right).
The diagrams are complicated, showing numerous avoided crossings and unstable strange modes; the extremely unstable mode around $\log (\Teff/{\rm K}) \approx 3.8$ appears to correspond to strange mode V found by \citep{1990b_Gautschy}\footnote{Note that these authors used a minimum modulus method with more success than shown in our Fig.~\ref{fig:0_9_modal_diag}; this could be a consequence of adopting a different discriminant function than \gyre.}.  All methods capture some aspects of this complexity, but only the contour method results in a complete modal diagram.


\section{Discussion} 
\label{sec:discussion}

In this paper we introduce the contour method as a new way of generating initial trial frequencies that can be used to find the complex roots of a discriminant function, $\Dw$, in the calculation of non-adiabatic stellar pulsations. The contour method involves evaluating the real and imaginary parts of $\Dw$ on a complex-$\omega$ grid, constructing the zero-contours for each part, and then searching for contour intersections to serve as initial trial frequencies (Sec.~\ref{sec:contour}).

We demonstrate the contour method implemented in the \gyre\ code by
calculating non-adiabatic pulsation frequencies for $10\,\Msun$ and $20\,\Msun$ $\beta$ Cephei star models, and for a 0.9$\Msun$ EHe star model (Sec.~\ref{sec:calculations}). Compared with the
adiabatic method and with the minimum modulus method
(Sec. \ref{sec:background}), the contour method finds all the
non-adiabatic pulsation frequencies within the given frequency range,
especially as the modes become more non-adiabatic.


The contour method is not entirely novel; \citet{1971_Dennis} used plots of the zero-contours of a discriminant function in the complex plane to explore thermal instabilities of $15\,\Msun$
He-shell burning models. However, the contour method has not been used for non-adiabatic pulsation problems, and it has not been automated using the marching-squares algorithm.

The contour method is also somewhat related to the method for finding initial trial frequencies described by \cite{1981_Shibahashi}. In their
method, they map closed loops in the complex-$\omega$ plane to the
complex-$\Dw$ plane. A loop winding around a root in the $\omega$
plane will wind around the origin in the $\Dw$ plane. If each loop is size of a single grid cell, and if $\Dw$ can locally be approximated as linear in $\omega$, then it can be shown that the two methods become equivalent. The contour method, however, has the additional benefit of
creating maps that visually display the global non-adiabatic pulsation properties of a model.


The main drawback to the contour method is its computational cost. For the $10\,\Msun$ snapshot considered in Sec.~\ref{sec:calculations},  the adiabatic method requires 31 seconds to calculate the modes shown in the bottom panel of Fig.~\ref{fig:10_0_ad_nad_map} (timings based on using a single core
of a 2.60GHz Intel E5-2690v4 processor). The minimum modulus method takes 102 seconds for the same calculation, and the contour method 300 minutes.

\begin{figure*}
	\centering
	\includegraphics[width=\linewidth,trim=0.1in 0.2in 0in 0.1in,clip]{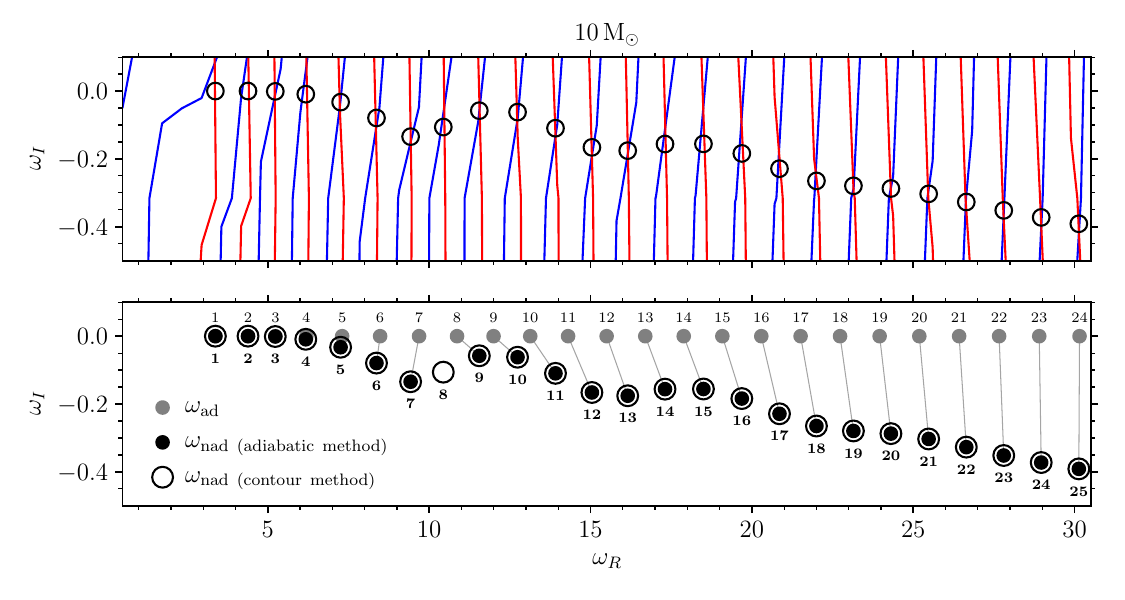}
	\caption{As in
            Fig.~\ref{fig:10_0_ad_nad_map}, except that 20 times fewer
            points in $\omegar$ and $\omegai$ are used in creating the
            contour map. Note how the contour intersections in the
            upper panel no longer precisely coincide with the
            non-adiabatic frequencies.}
	\label{fig:10_0_ad_nad_map_x20}
\end{figure*}

The expensive part of the contour method is evaluating $\Dr$ and $\Di$
at every point on the grid. However, this expense can be mitigated in
two ways. The first is that the evaluations are embarrassingly
parallel and can take advantage of multiple cores and/or cluster nodes. Distributing the calculations across 28 E5-2690v4 cores via Message Passing Interface (MPI) calls reduces the calculation time of the contour method to 12 minutes, a nearly linear speed-up. The second is that the contour method remains viable with a low resolution grid; the contour map and the resulting intersections will be less accurate but the intersections can still serve as sufficiently accurate initial trial frequencies.  In Fig.~\ref{fig:10_0_ad_nad_map_x20} we show the contour map and
non-adiabatic pulsation frequencies found for the $10\,\Msun$
snapshot, with 20 times fewer points than previously in both $\omegar$ and $\omegai$. The contour map is jagged and the pulsation frequencies found are no longer centered on the intersections, as in Fig. \ref{fig:10_0_ad_nad_map}, but the contour method still finds all non-adiabatic frequencies. With this grid resolution, the calculation takes 66 seconds on a single core, around 270 times faster than the original run.

The computational expense of the contour method means that it isn't always the best approach to calculating pulsation frequencies. When non-adiabatic effects are very weak --- for instance, in slowly pulsating B stars, whose gravity modes are characterized by small growth/damping rates --- the adiabatic method for obtaining initial trial frequencies remains sufficient. However, for stars with larger growth rates such as the $\beta$ Cephei and EHe stars modeled here, together with other pulsators at high luminosity-to-mass ratios, the contour method succeeds when the adiabatic method fails.

This is particularly relevant now as we expect unprecedented pulsation data from \emph{TESS} \citep{2014_Ricker}, which will observe the variability of high-luminosity stars previously excluded in directed campaigns (e.g., \emph{Kepler}). These data, analyzed using the contour method, can be applied to model and test non-adiabatic pulsations across the HRD, providing fresh insights into stellar structure and evolution.


\pagebreak

\section{Acknowledgments} \label{sec:acknowledgements}

We thank Ellen Zweibel and Lars Bildsten for their insightful input during the project, and the anonymous referee for their helpful remarks. We acknowledge support from National Science Foundation grants ACI-1663696, AST-1716436 and PHY-1748958. This
research was performed using the compute resources and assistance of
the UW-Madison Center For High Throughput Computing (CHTC) in the
Department of Computer Sciences. The CHTC is supported by UW-Madison,
the Advanced Computing Initiative, the Wisconsin Alumni Research
Foundation, the Wisconsin Institutes for Discovery, and the National
Science Foundation, and is an active member of the Open Science Grid,
which is supported by the National Science Foundation and the
U.S. Department of Energy's Office of Science.


\pagebreak

\bibliographystyle{aasjournal}
\bibliography{contour}

\begin{thebibliography}{}
\expandafter\ifx\csname natexlab\endcsname\relax\def\natexlab#1{#1}\fi
\providecommand{\url}[1]{\href{#1}{#1}}

\bibitem[{{Aerts} {et~al.}(2010){Aerts}, {Christensen-Dalsgaard}, \&
  {Kurtz}}]{2010_Aerts_etal}
{Aerts}, C., {Christensen-Dalsgaard}, J., \& {Kurtz}, D.~W. 2010,
  {Asteroseismology} (Springer, Netherlands)

\bibitem[{{Asplund} {et~al.}(2009){Asplund}, {Grevesse}, {Sauval}, \&
  {Scott}}]{2009_Asplund}
{Asplund}, M., {Grevesse}, N., {Sauval}, A.~J., \& {Scott}, P. 2009, \araa, 47,
  481

\bibitem[{{Castor}(1971)}]{1971_Castor}
{Castor}, J.~I. 1971, \apj, 166, 109

\bibitem[{{Cox} {et~al.}(1992){Cox}, {Morgan}, {Rogers}, \&
  {Iglesias}}]{1992_Cox}
{Cox}, A.~N., {Morgan}, S.~M., {Rogers}, F.~J., \& {Iglesias}, C.~A. 1992,
  \apj, 393, 272

\bibitem[{{Dennis}(1971)}]{1971_Dennis}
{Dennis}, T.~R. 1971, \apj, 167, 311

\bibitem[{{Dziembowski} \& {Pamiatnykh}(1993)}]{1993a_Dziembowski}
{Dziembowski}, W.~A., \& {Pamiatnykh}, A.~A. 1993, \mnras, 262, 204

\bibitem[{{Gautschy} \& {Glatzel}(1990{\natexlab{a}})}]{1990a_Gautschy}
{Gautschy}, A., \& {Glatzel}, W. 1990{\natexlab{a}}, \mnras, 245, 154

\bibitem[{{Gautschy} \& {Glatzel}(1990{\natexlab{b}})}]{1990b_Gautschy}
---. 1990{\natexlab{b}}, \mnras, 245, 597

\bibitem[{{Jeffery}(2008)}]{2008a_Jeffery}
{Jeffery}, C.~S. 2008, in ASP Conf. Ser. 391: Hydrogen-Deficient Stars, ed.
  A.~{Werner} \& T.~{Rauch}, 53

\bibitem[{{Ledoux} \& {Walraven}(1958)}]{1958_Ledoux}
{Ledoux}, P., \& {Walraven}, T. 1958, Handbuch der Physik, 51, 353

\bibitem[{{Paxton} {et~al.}(2011){Paxton}, {Bildsten}, {Dotter}, {Herwig},
  {Lesaffre}, \& {Timmes}}]{2011_Paxton}
{Paxton}, B., {Bildsten}, L., {Dotter}, A., {et~al.} 2011, \apjs, 192, 3

\bibitem[{{Paxton} {et~al.}(2013){Paxton}, {Cantiello}, {Arras}, {Bildsten},
  {Brown}, {Dotter}, {Mankovich}, {Montgomery}, {Stello}, {Timmes}, \&
  {Townsend}}]{2013_Paxton}
{Paxton}, B., {Cantiello}, M., {Arras}, P., {et~al.} 2013, \apjs, 208, 4

\bibitem[{{Paxton} {et~al.}(2015){Paxton}, {Marchant}, {Schwab}, {Bauer},
  {Bildsten}, {Cantiello}, {Dessart}, {Farmer}, {Hu}, {Langer}, {Townsend},
  {Townsley}, \& {Timmes}}]{2015_Paxton}
{Paxton}, B., {Marchant}, P., {Schwab}, J., {et~al.} 2015, \apjs, 220, 15

\bibitem[{{Paxton} {et~al.}(2018){Paxton}, {Schwab}, {Bauer}, {Bildsten},
  {Blinnikov}, {Duffell}, {Farmer}, {Goldberg}, {Marchant}, {Sorokina},
  {Thoul}, {Townsend}, \& {Timmes}}]{2018_Paxton}
{Paxton}, B., {Schwab}, J., {Bauer}, E.~B., {et~al.} 2018, \apjs, 234, 34

\bibitem[{{Paxton} {et~al.}(2019){Paxton}, {Smolec}, {Gautschy}, {Bildsten},
  {Cantiello}, {Dotter}, {Farmer}, {Goldberg}, {Jermyn}, {Kanbur}, {Marchant},
  {Schwab}, {Thoul}, {Townsend}, {Wolf}, {Zhang}, \& {Timmes}}]{2019_Paxton}
{Paxton}, B., {Smolec}, R., {Gautschy}, A., {et~al.} 2019, \apjs, 243, 10

\bibitem[{{Press} {et~al.}(1992){Press}, {Teukolsky}, {Vetterling}, \&
  {Flannery}}]{1992_Press}
{Press}, W.~H., {Teukolsky}, S.~A., {Vetterling}, W.~T., \& {Flannery}, B.~P.
  1992, Numerical recipes in Fortran 77, 2nd edn. (University of Cambridge
  Press, Cambridge)

\bibitem[{{Ricker} {et~al.}(2014){Ricker}, {Winn}, {Vanderspek}, {Latham},
  {Bakos}, {Bean}, {Berta-Thompson}, {Brown}, {Buchhave}, {Butler}, {Butler},
  {Chaplin}, {Charbonneau}, {Christensen-Dalsgaard}, {Clampin}, {Deming},
  {Doty}, {De Lee}, {Dressing}, {Dunham}, {Endl}, {Fressin}, {Ge}, {Henning},
  {Holman}, {Howard}, {Ida}, {Jenkins}, {Jernigan}, {Johnson}, {Kaltenegger},
  {Kawai}, {Kjeldsen}, {Laughlin}, {Levine}, {Lin}, {Lissauer}, {MacQueen},
  {Marcy}, {McCullough}, {Morton}, {Narita}, {Paegert}, {Palle}, {Pepe},
  {Pepper}, {Quirrenbach}, {Rinehart}, {Sasselov}, {Sato}, {Seager},
  {Sozzetti}, {Stassun}, {Sullivan}, {Szentgyorgyi}, {Torres}, {Udry}, \&
  {Villasenor}}]{2014_Ricker}
{Ricker}, G.~R., {Winn}, J.~N., {Vanderspek}, R., {et~al.} 2014, in Proc. SPIE,
  Vol. 9143, Space Telescopes and Instrumentation 2014: Optical, Infrared, and
  Millimeter Wave, 914320

\bibitem[{{Saio} {et~al.}(1984){Saio}, {Wheeler}, \& {Cox}}]{1984_Saio}
{Saio}, H., {Wheeler}, J.~C., \& {Cox}, J.~P. 1984, \apj, 281, 318

\bibitem[{{Shibahashi} \& {Osaki}(1981)}]{1981_Shibahashi}
{Shibahashi}, H., \& {Osaki}, Y. 1981, \pasj, 33, 427

\bibitem[{{Stankov} \& {Handler}(2005)}]{2005_Stankov}
{Stankov}, A., \& {Handler}, G. 2005, \apjs, 158, 193

\bibitem[{{Suran}(2008)}]{2008_Suran}
{Suran}, M.~D. 2008, \apss, 316, 163

\bibitem[{{Townsend}(2005)}]{2005a_Townsend}
{Townsend}, R.~H.~D. 2005, \mnras, 360, 465

\bibitem[{{Townsend} {et~al.}(2018){Townsend}, {Goldstein}, \&
  {Zweibel}}]{2018_Townsend}
{Townsend}, R.~H.~D., {Goldstein}, J., \& {Zweibel}, E.~G. 2018, \mnras, 475,
  879

\bibitem[{{Townsend} \& {Teitler}(2013)}]{2013_Townsend}
{Townsend}, R.~H.~D., \& {Teitler}, S.~A. 2013, \mnras, 435, 3406

\bibitem[{{Unno} {et~al.}(1989){Unno}, {Osaki}, {Ando}, {Saio}, \&
  {Shibahashi}}]{1989_Unno}
{Unno}, W., {Osaki}, Y., {Ando}, H., {Saio}, H., \& {Shibahashi}, H. 1989,
  {Nonradial Oscillations of Stars}, 2nd edn. (University of Tokyo Press,
  Tokyo)

\bibitem[{{Weiss}(1987)}]{1987_Weiss}
{Weiss}, A. 1987, \aap, 185, 178

\bibitem[{Wenger(2013)}]{2013_Wenger}
Wenger, R. 2013, Isosurfaces: Geometry, Toplogy, and Algorithms (Tsylor \&
  Francis Group, Boca Raton)

\end{thebibliography}

\end{document}